\documentclass{elsarticle}

\usepackage{paralist}
\usepackage{graphicx,floatrow}
\usepackage{graphicx}
\usepackage{algorithm}
\usepackage[noend]{algorithmic}
\usepackage{color}
\usepackage{amsmath,amssymb}
\usepackage{amstext}
\usepackage{appendix}
\usepackage{subfigure}
\usepackage{multirow}
\usepackage{balance,url}
\usepackage{mdwlist}
\usepackage{balance}
\usepackage{indentfirst}
\usepackage{hyperref}
\usepackage{array}
\usepackage{subfigure}
\usepackage{lineno,hyperref}
\modulolinenumbers[5]

\begin{document}

\begin{frontmatter}

\title{GPU-based Efficient Join Algorithms on Hadoop}

\author{Hongzhi Wang, Ning Li , Zheng Wang, Jianing Li}
\address{P.O.Box 750, Harbin Institute of Technology, Harbin, China}

\begin{abstract}
\par The growing data has brought tremendous pressure for query processing and storage, so there are many studies that focus on using GPU to accelerate join operation, which is one of the most important operations in modern database systems. However, existing GPU acceleration join operation researches are not very suitable for the join operation on big data.
\par Based on this, this paper speeds up nested loop join, hash join and theta join, combining Hadoop with GPU, which is also the first to use GPU to accelerate theta join. At the same time, after the data pre-filtering and pre-processing, using Map-Reduce and HDFS in Hadoop proposed in this paper, the larger data table can be handled, compared to existing GPU acceleration methods. Also with Map-Reduce in Hadoop, the algorithm proposed in this paper can estimate the number of results more accurately and allocate the appropriate storage space without unnecessary costs, making it more efficient.
\par The rigorous experiments show that the proposed method can obtain 1.5 to 2 times the speedup, compared to the traditional GPU acceleration equi join algorithm. And in the synthetic data set, the GPU version of the proposed method can get 1.3 to 2 times the speedup, compared to CPU version.
\end{abstract}

\end{frontmatter}


\section{Introduction}
\par One of the most serious problems in the computer industry today is the growing data. According to statistics, the rate at which data is generated annually on the network will increase by 10 percent every five years\cite{Gantz}. Therefore, we have to face how to effectively deal with this serious problem in large relational databases. However, the speed of the processor has now grown to the limits of the current level of technology, more and more attention focused on parallel technology. One solution is to increase the number of processors\cite{Krueger} and threads\cite{Low}. Another solution is using a single instruction multiple data stream (SIMD) structure to improve the parallelism by processing multiple data under one instruction\cite{Zhou}.
\par Now, due to CPU clock frequency limitations, software optimization has come to an end. Therefore, researchers have to consider other possibilities to speed up the query processing\cite{ZHANG2017}, using multicore CPUs\cite{SILVA2017402}. The use of new hardware to speed up the process is also possible\cite{David}. In recent years, many studies have chosen FPGA as an option for hardware accelerators\cite{Teubner, Woods}, including the query\cite{SINGARAJU2015149}. Similarly, image processors (GPU) have been widely used in the field of query processing\cite{HERNANDEZ2017}.
\par Join operation is one of the most important operations in relational database operations and one of the longest-running operations in a query. Under this situation, there are many efforts put to speed up the join operation. IBM has added new hardware to its commercial system Netezza\cite{Netezza}. Do\cite{Do} integrated CPU processors and DRAM memory into a smart flash device (Smart SSD) to implement query processing. Devarajan\cite{Devarajan} and others believe that the GPU is the most advanced distributed tool to handle computationally intensive tasks. Kaldewey\cite{Kaldewey} and others believed the data needs to be copied to the GPU device memory for processing. He\cite{He} thought using GPU to query co-processing was an effective way to improve memory database performance. Yuan\cite{Yuan} used the GPU device to implement the hash join operation. Pietron\cite{Pietron} and others used the CUDA programming model on the GPU to achieve a part of SQL operations. Rui\cite{Rui} has conducted detailed experimental studies on how join operation benefits from the rapid growth of the GPU. Angstadt\cite{Angstadt} even developed a dedicated to speed up the SQL statement using the CUDA programming model on the GPU. In addition to the two-table join, multi-table join equally occupies a very important position in the relational database, Zhou\cite{Zhou1}, who proposed GBFSJ (GPUs BloomFilter Star Join) algorithm,  achieved a star join on the GPU with the use of a Bloom filter. Cruz\cite{Cruz} and others have used the GPU to achieve such connectivity.
\par Through the analysis of the existing research results of the new hardware acceleration join, it is concluded that the research results have the following problems:
\begin{itemize}
	\item Existing join operations based on new hardware are still at the initial stage, most are limited to simple equi join, or lack of research on complex join operation such as theta join. Future research work should focus more on complex join operation for practical applications.
	\item The existing research carried out by the experiment are based on small data sets, the size of the data set mostly is MB. To use GPU in commercial database systems, future research efforts should be put on scaling operations on large-scale data.
	\item A serious problem with the new hardware acceleration join operation is how to allocate the appropriate storage space for the join results. Different from the CPU programming language, dynamically allocating storage space, GPU needs to allocate enough storage space in advance. However, the existing research results are not good solutions to this problem.
	\item A single new hardware cannot meet the needs of modern business databases, but existing research is carried out on monolithic new hardware. There should be more work on deploying new hardware as a distributed architecture to accelerate join operation.
\end{itemize}
\par This paper is mainly based on the distributed architecture of the GPU to accelerate the join operation. By combining the Hadoop architecture with the GPU, advantages of both Hadoop's node-level parallelism and GPU's thread-level parallelism can be taken. Hadoop data processing tasks originally performed by the CPU were sent to the GPU, using the GPU's parallelism while opening multiple threads to take advantage of the high computational power and high parallelism. This paper intends to implement nested loop join, hash join and theta join algorithm, the remaining types of join will be studied later.
\par  The second section mainly introduces the basic hardware structure, the thread organization form and the CUDA programming language background, to better understand the following GPU processing join algorithm. The third section and the fourth section mainly introduce the main research contents of this paper, including pre-filtering of data and hardware processing equi join and non-equi join operation. The fifth section introduces the experiment and the results obtained in this paper. The last part concludes the paper, summarizes the contents and points out the innovation.
\section{Preliminary}
\subsection{GPU}
\par
\begin{figure}[h]
	\centering\includegraphics[width=7cm, height=8.5cm]{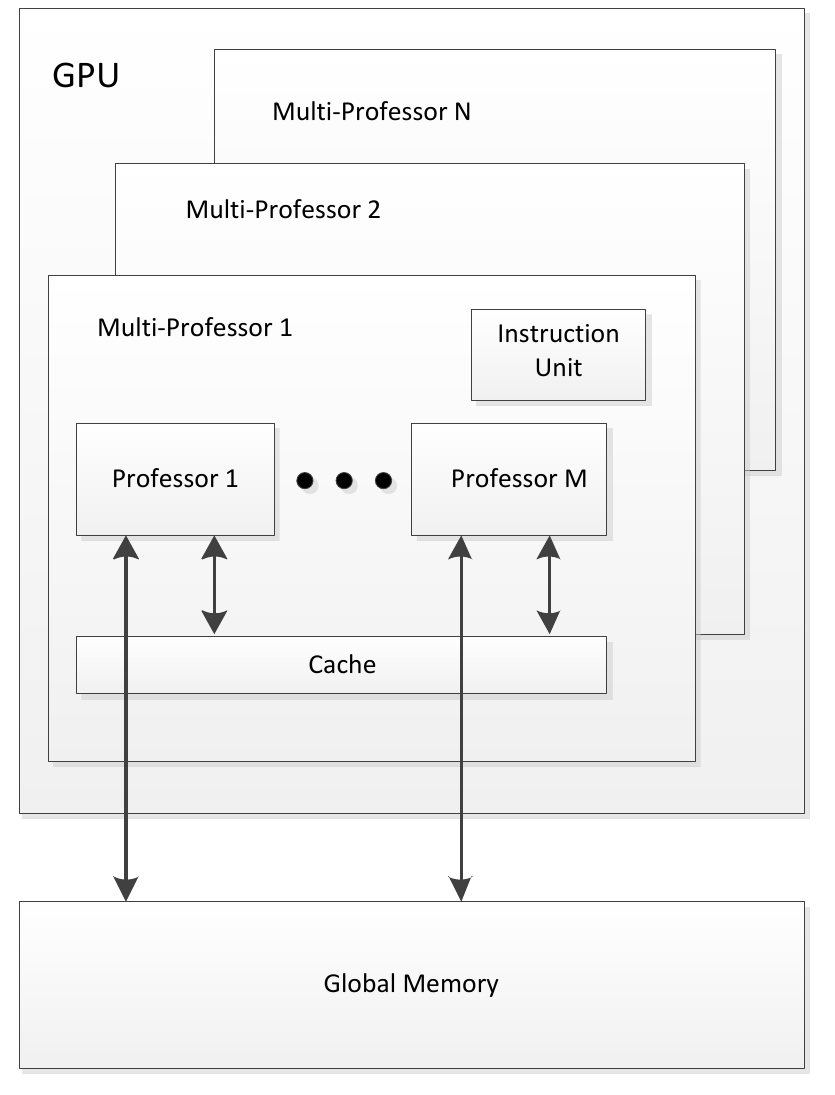}
	\caption{Structure of GPU}\label{GPU1}
\end{figure}

 The GPU device has a multiparty processor core with multiple instruction streams and multiple streams (SIMDs), and each multiprocessor core contains a number of processors. The GPU hardware structure shown in Figure~\ref{GPU1} contains $N$ multiprocessor cores,
and each multiprocessor core contains $M$ processors. Each multiprocessor core contains an instruction processing unit and a storage resource. All processors on the multiprocessor share the instruction unit and the storage space, and each processor has a set of registers. GPU devices also have global memory, and global memory can be accessed by all multiprocessor cores. Multi-processor core internal storage resources read and write faster, compared to the global memory.
\par
\begin{figure}[h]
	\centering\includegraphics[width=9cm, height=0.8\textwidth]{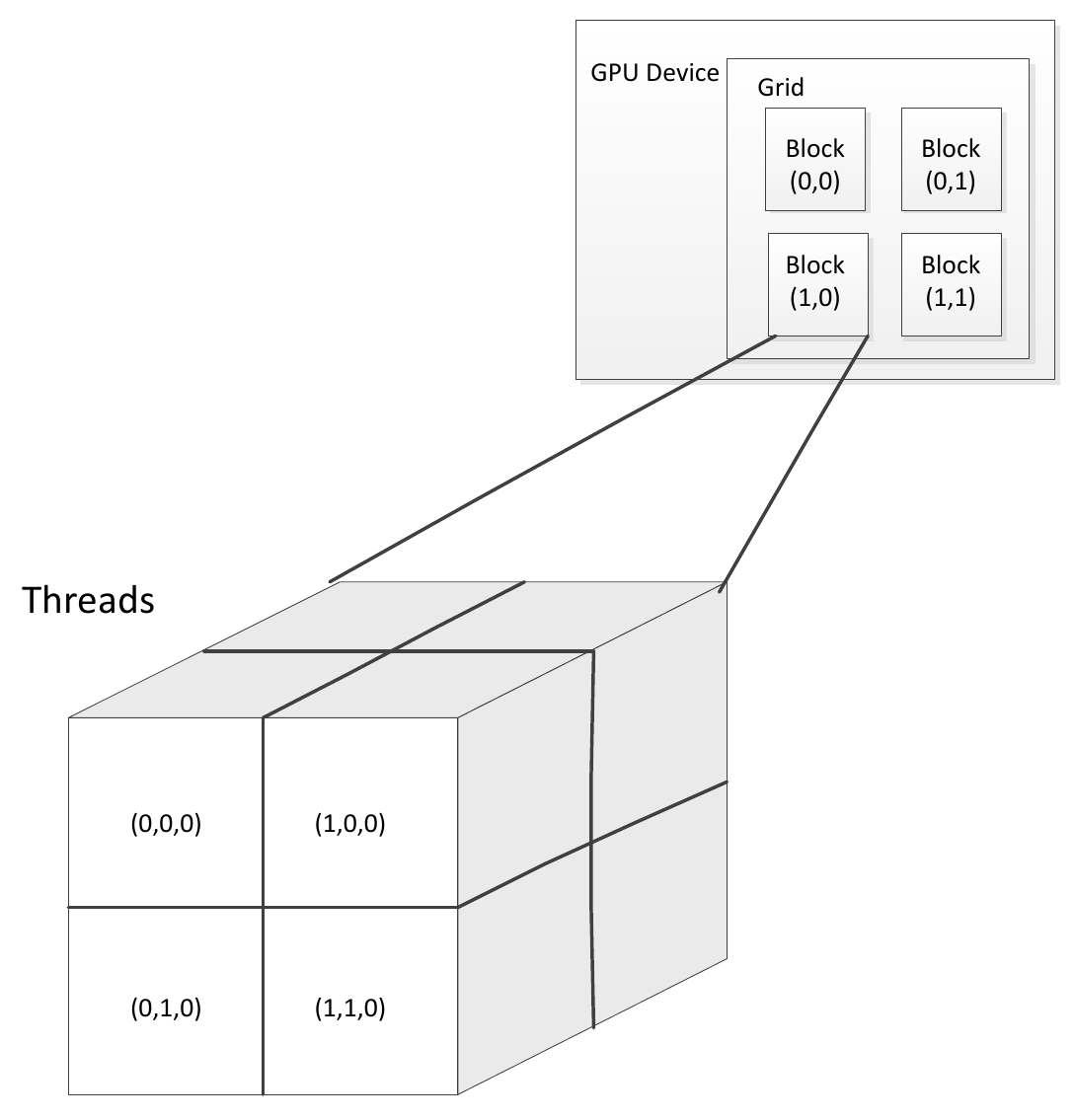}
	\caption{Organization of GPU threads}\label{GPU2}
\end{figure}

 As shown in Figure~\ref{GPU2}, in a thread block, the organization of the thread can be one-dimensional, two-dimensional and three-dimensional, with ID identifying each thread. 
Similarly, the organization of thread blocks in a thread grid can also be divided into one-dimensional, two-dimensional, and three-dimensional. GPU threads have a variety of organizational forms, making it applicable to different issues. For example, when dealing with array problems, the thread grid and the thread block can be used in the one-dimensional organization, and each thread corresponds to a part of the array elements. \subsection{CUDA}\label{CUDA}
\par GPU devices slowly evolved into a copier processor for intensive computation, bringing a variety of programming languages into birth, such as CUDA. CUDA can be implemented in both CPU and GPU. Additionally, CUDA adds some content related to GPU devices, such as the use of rich thread resources on the GPU.
\par CUDA contains two kinds of code, host code, and kernel code. Host code runs on the CPU, which is responsible for the applying for storage space,  calling the kernel code, controlling data transformation between CPU and GPU. The kernel code is the code that runs in parallel on the GPU.
\subsection{Join Operation}
\par In the relational database, the join operation is the process of the combination of two tables into a relationship table under specific conditions. The attributes that participate in the relationship table are called join keys. If the join key satisfies the query condition, the corresponding tuples in two tables are merged into one tuple and stored in the buffer.
\par According to the different join conditions, the join operation can be divided into equi join and theta join:
\begin{itemize}
	\item Equi join: The query statement specifies the join condition for the connection of the equation. Consider the relationships $R(A,B)$, $S(C,D)$. When $R.A=S.C$, it is an equi join. Figure~\ref{equi_join} shows the result of the join. In SQL, the syntax of this join is:
	\par $select$ $A,B,C,D$ $from$ $R$ $join$ $S$ $on$ $R.A = S.C$
	\begin{figure}[h]
		\centering\includegraphics[width=4.1in]{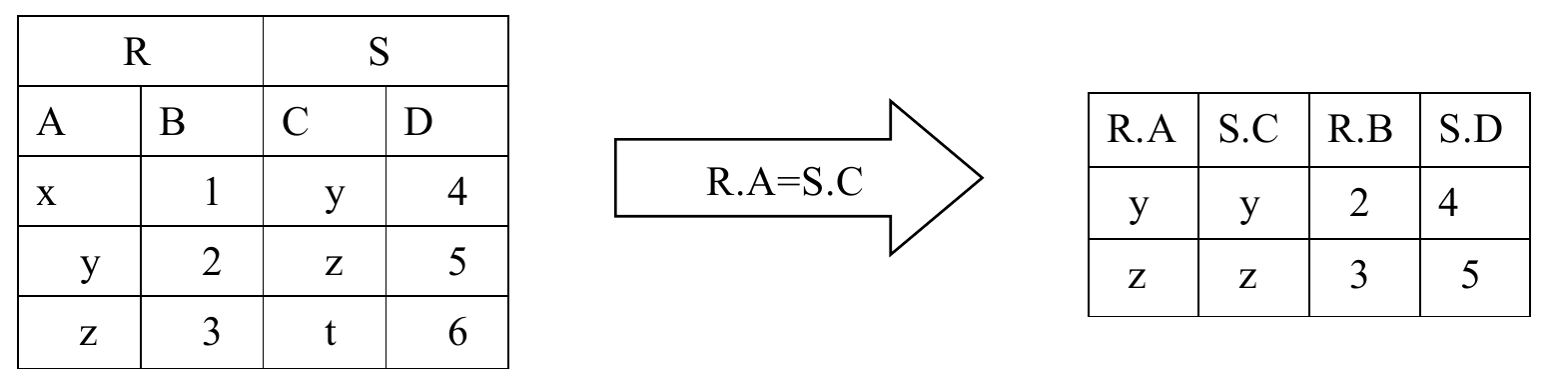}
		\caption{Example of equi join}\label{equi_join}
	\end{figure}
	\item Theta join: The query statement specifies the join condition for the connection of the non-equation. Non-equation includes $>,<,\ge,\le$ and so on. Figure~\ref{nonequi_join} shows the result of the join. In SQL, the syntax of this join is:
	\par $select$ $A,B,C,D$ $from$ $R$ $join$ $S$ $on$ $R.A < S.C$
	\begin{figure}[h]
		\centering\includegraphics[width=4.1in]{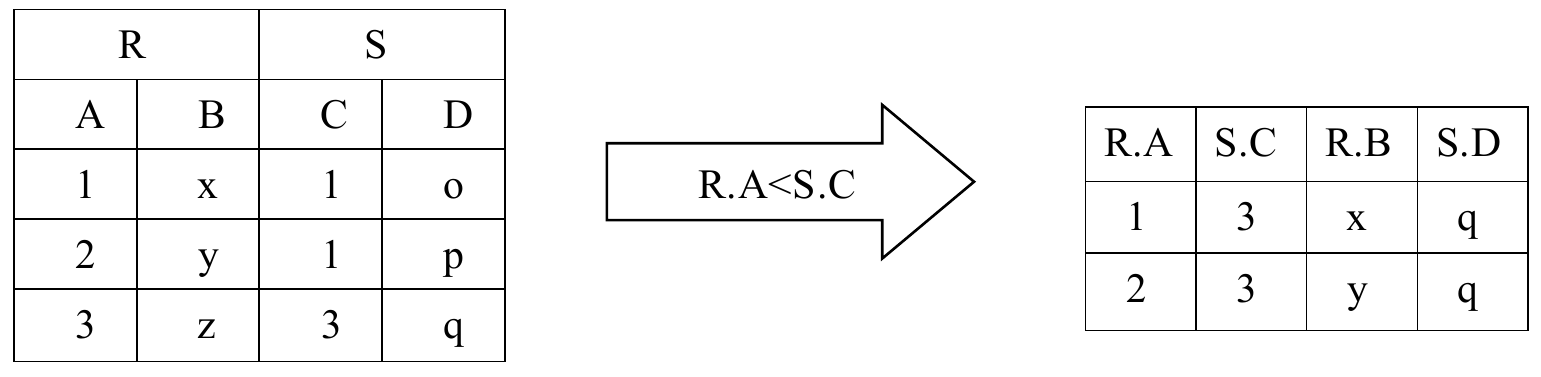}
		\caption{Example of theta join}\label{nonequi_join}
	\end{figure}
\end{itemize}
The join operation can be roughly divided into the following three categories:
\begin{itemize}
	\item Nested loop join\label{nestedloop}: It is a violent algorithm that converts all tuples in one table to all tuples in the other table, and generates a new result tuple if the join condition is met.
	\item Hash join: It is to put the smaller table (inner table) into a hash table and store it in memory. And then traverse the larger table (outer table) to find the tuple of the outer table in the hash table.
	\item Sort merge join: It is to sort two tables at first, and then traverse them in sequence to decide whether to join.
\end{itemize}
\subsection{Hadoop}
Hadoop is an open source software framework, which is used for distributed storage and big data processing. The core of this framework is the Hadoop Distributed File System (HDFS) and Map-Reduce.
\par HDFS, written in Java for the Hadoop framework, is a distributed file system that can store data on commodity machines, providing high bandwidth across the cluster.
\par Map-Reduce is a distributed computing framework to handle big data (greater than 1TB), including Map and Reduce. In the Map phase, the data is read in units of lines and converted to $key-value$ pairs. In the Reduce phase, pairs with the same $key$ value are passed to the same Reduce function, and the result is finally returned\cite{Myung, Tong, Mohamad, Theresa, Foto}.
\par Using Map-Reduce to handle join operations is the current mainstream approach, roughly divided into three ways, join in Reduce, join in Map, and semi join.
\section{GPU-Based Equi Join with Hadoop}\label{gpubej}
\par In the join algorithm above, only a small part of the data in two tables participates in the equi join operation. Therefore, initial pre-filtering in CPU, passing fewer data between hosts, will be a good solution to this problem. Based on this, this section mainly introduces GPU to accelerate nested loop join and hash join, along with the pre-filtering through Hadoop to reduce the amount of data transferred between devices.
\subsection{Data Pre-filtering through Hadoop}
\par In semi join, only tuples of the large group are filtered, by extracting the join key of the small table. When two large tables are connected, filtering only the larger ones is not sufficient to achieve the greatest performance boost. The pre-filter mentioned in this paper extracts common join keys of two tables to pre-filter both tables. So, even if two tables are very large, it can still achieve a good performance. This pre-filtering needs two rounds of Map-Reduce through Hadoop, the following is a specific implementation process.
\par In the Map phase of the first round of Map-Reduce, firstly read the data of two tables to extract those to be connected attributes, and add a label to indicate which table they are from. Specifically, Map's output is $(key, tag)$, and $key$ is the value of to be connected attribute. When $tag$ is T, it means the tuple is from the table T, the same to those with S. Each tuple corresponds to such a $key-value$ pair. Pairs with the same $key$ are transported to the same Reducer after shuffling, sorting, merging, and so on. In the Reduce phase, the received $(key, value_list)$ are analyzed. Only those $value_list$ contains both T and S key value will be outputted as the result to HDFS in Hadoop. All in all, the first round Map-Reduce's output is the value of join key attribute in the final result. That is, the tuples whose value of join key attribute are not in it will not appear in the final result.
\par In the second round of the Map-Reduce task, the result from the previous round is read from the HDFS firstly in the initialization function Setup() and stored in a hash table. Then in the Map phase, read tuples of two to be joined original table, and extract the join key. If the join key exists in the hash table, the tuple is shown in the final result, it is outputted with a tag that identifies its origin. In this step, only those tuples that are determined to be in the final result will be outputted to Reduce to participate in the final operation. In this way, the amount of data to be processed in Reduce declines greatly. At the same time, it reduces the number of tuples that need to be stored on the GPU, freeing up device memory.
\par In semi join, the raw data is inputted directly to HDFS after filtering, and then a new Map-Reduce is restarted to read and operate the filtered data. However, it will lead to additional Map-Reduce startup time and additional expense of both output and extraction in the new round. Therefore, the pre-filtering method used in this paper incorporates the filtering of data and the execution of the actual join operation into one round of tasks. That is, after filtering, mapping and tagging in Map, data is transferred to Reduce for joining, by shuffling, sorting, merging and other steps. In the Reduce phase, the data from Map is received and processed (row and column transformation), and the processed data is then sent to the GPU for specific join operations. After the GPU is executed, the result is returned to the CPU, and the Reduce outputs it to HDFS. The pseudo code is shown in Algorithm~\ref{prefilteringalgo}.
\begin{algorithm}[ht]
	\scriptsize
	\caption{Pre-filtering Algorithm}
	\label{prefilteringalgo}
	\begin{algorithmic}[1]
		\STATE Map1(null,$tuple$):
		\STATE $join\_key \leftarrow$ extract the join key from $tuple$;
		\STATE emit(join\_key,$tag$);
		\STATE Reduce1($key$,$tag\_list$):
		\STATE $unique\_key \leftarrow$ the key which belongs to both of the two tables;
		\FOR {key in $unique\_key\_list$}
		\STATE emit($key$,null);
		\ENDFOR
		\STATE Setup()
		\STATE Build a hash table with the $unique\_key\_list$;
		\STATE Map2(null,$tuple$):
		\STATE $join\_key \leftarrow$ extract the join key from $tuple$;
		\STATE $join\_tuple \leftarrow$ $tuple$ tuple whose join key is contained in the hash table;
		\STATE $join\_tuple^, \leftarrow$ $Projection(Join\_tuple)$;
		\STATE emit($join\_key/a$,tagged join\_tuple);
		\STATE Reduce2($key$,$tag\_list$):
		\STATE $T$ $\leftarrow$ tuples from table T for $key$;
		\STATE $S$ $\leftarrow$ tuples from table S for $key$;
		\STATE $T^,$ $\leftarrow$ preprocessing for table $T^,$;
		\STATE $S^,$ $\leftarrow$ preprocessing for table $S^,$;
		\STATE $NB\_T$ $\leftarrow$ number of tuples in $T^,$;
		\STATE $NB\_S$ $\leftarrow$ number of tuples in $S^,$;
		\STATE $Result$ $\leftarrow$ $Join\_GPU(T^,,S^,,NB\_T, NB\_S)$;
		\FOR {$tuple$ in $Result$}
		\STATE emit(null,$tuple$);
		\ENDFOR
	\end{algorithmic}
\end{algorithm}
\subsection{Data Preprocessing through Hadoop}
Before reading the filtered data and performing the actual join operation, Reduce of Map-Reduce in Hadoop needs to perform some preprocessing of the data. The main processing steps include mapping and row and column transformation. This section describes these two operations and the effects they bring.
\subsubsection{Mapping}
\par For a SQL query, only a few attributes in the relational table will be used, and most of them will lead to additional overhead. Therefore, analyzing SQL statement to determine which attributes are query-related will enhance the efficiency. Existing big data query tools, such as HIVE, IMPALA, all follow a principle: by analyzing the SQL statement, generate the implementation plan, and optimize the implementation plan, including filtering, mapping operations before the join operation.
The algorithm presented in this paper will map the tuples in the Map phase of the second round of Map-Reduce. For each tuple, only query-related attributes are kept, others are removed. Those kept attributes are combined into new tuples. The new tuple has all the query-related attribute values, but the size is smaller than the original tuple. The pseudo code is shown in Algorithm~\ref{mappingalgo}.
\begin{algorithm}[ht]
	\scriptsize
	\caption{Mapping Algorithm(Projection)}
	\label{mappingalgo}
	\begin{algorithmic}[1]
		\STATE $new\_tuple \leftarrow$ null;
		\STATE $Attrs[] \leftarrow$ extract attributes from tuple T;
		\STATE $new\_tuple \leftarrow$ combine relevant attributes in $Attrs[]$ to a new tuple;
	\end{algorithmic}
\end{algorithm}
\subsubsection{Row and Column Transformation}
\par Traditional relational databases use row storage patterns, using tuple as a basic storage unit in the disk. All bytes of each tuple are stored adjacently. In online transaction processing (OLTP), the query needs to return all or most of the columns, making this storage mode perform well.
\par However, in online analytical processing (OLAP), queries often involve millions or more row tuples, but only a few of them are required for queries. For example, a supermarket needs to check the top ten highest sales of commodities this year. Users usually only concern about the name, categories, and sales. At this point, the using of row storage will read a lot of useless data, because only one line of data can be read at one time, of which only a few field value is required for the query.
\par
\begin{figure}[h]
	\centering\includegraphics[width=5in]{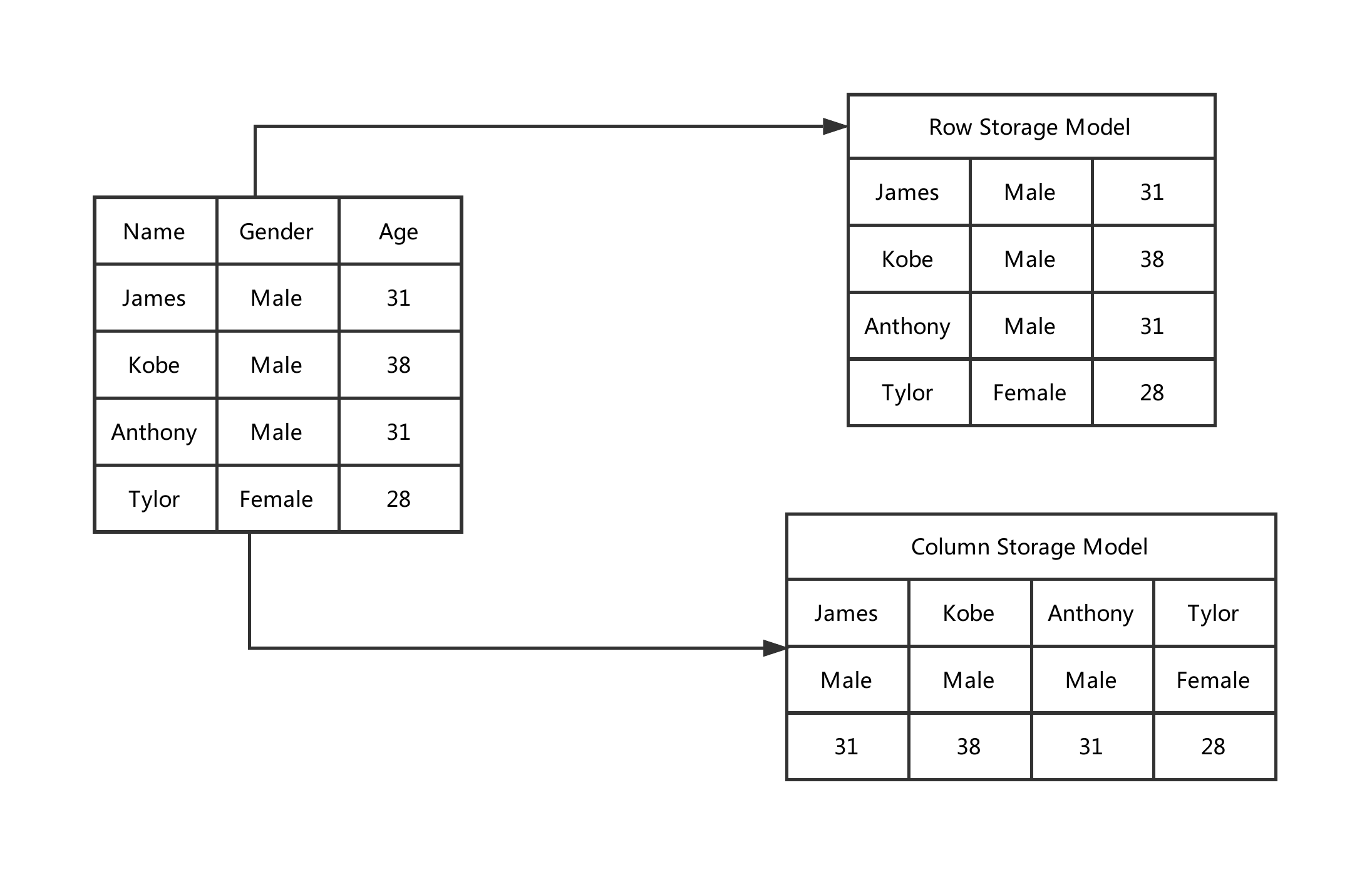}
	\caption{Column storage model}\label{columnstore}
\end{figure}

The column storage model is a good solution to this flaw. As shown in Figure~\ref{columnstore}, the column storage model stores the data in columns, and all the data in the same column are stored in adjacent storage space. When storing one row of data, different data fields are stored in the corresponding column's storage space. For the column storage model, it only needs to read the columns associated with the query. In addition, the column storage model performs better in data compression than the line storage model. Although the column storage model has many advantages, after being read from the disk, data still needs to be stored in rows in memory for processing. When the number of to be read columns is too large, the process of row and column transformation will bring too much overhead. Therefore, the column storage model is more suitable for the query with fewer columns involved. At present, many database systems use column storage mode, such as SQL Server\cite{Larson}.
\par In this paper, because the data has been mapped in the Map of the second round, row and column conversion cannot bring too many benefits. However, for the join operation, a necessary step is to extract the join key value of each tuple. As is shown in Chapter~\ref{CUDA}, CUDA is not quite good at the string processing, for the need of traversing the entire tuple to extract the join key, which will undoubtedly increase the computing burden of the GPU. At the same time, because the tuple contains variables of unsettled length, extracting the join key is not a very wise choice in the kernel. So before sending data to the GPU, row and column still need to be transformed.
\par To do the transformation, extract all the attribute values of each tuple and store these attribute values into different buffers. Each buffer stores all the values of an attribute. Note that attributes in each buffer are stored in the same order. For any two different tuples t and s in the original data table, let tuple t be preceded by tuple s. The position of each attribute of the tuple t in the respective buffer is preceded by the corresponding attribute of the tuple. As a result, only comparing the records in the join key buffer and extracting the records in other locations to generate join results are needed. For example, if the $m^{th}$ record in the T table join key buffer matches the $n^{th}$ record in the S table join key buffer, extract the joining result of all $m^{th}$ records of T table buffers and all $n^{th}$ records of S table buffers.
\subsection{GPU-Accelerated Equi Join Operation}
\par This section describes the details of equi join operations on the GPU, including the threading model used in CUDA, the specific execution flow of join operations, the estimation and analysis of results number, and the thread hierarchy analysis. In addition, this section implements nested loop joins and hash join operations.
\subsubsection{GPU-Based Nested Loop Join}
\par
\begin{figure}[h]
	\centering\includegraphics[width=0.4\textwidth,height=0.4\textwidth]{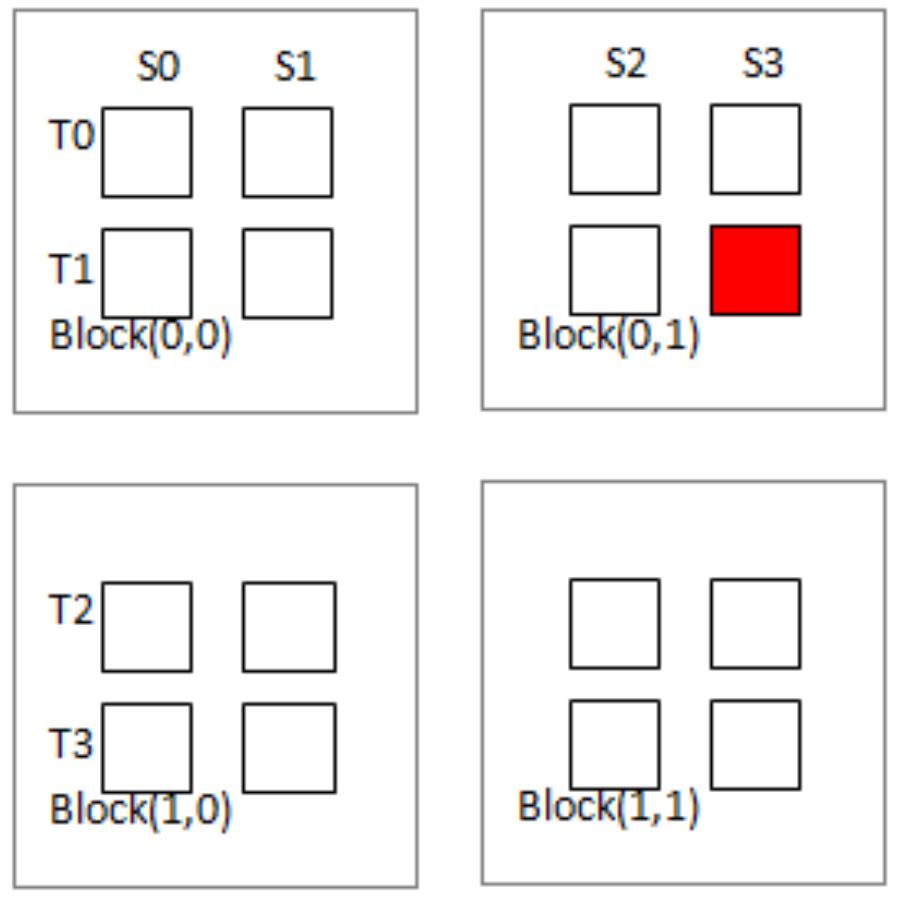}
	\caption{Thread grid organization of nested loop join}\label{nestedloopfig}
\end{figure}

 According to Section~\ref{nestedloop}, nested loop join needs to loop through two tables, so the use of two-dimensional thread model is essential, that is, the organization of the thread block in the thread grid and the organization of the thread in the thread block are two-dimensional. Figure~\ref{nestedloopfig} is a thread grid organization, which contains a total of four thread blocks, each thread block also contains four threads, and are in the form of $2*2$ organization. Thus, the line grid in both horizontal axis and vertical axis contain two thread blocks, the same as the organization of threads in the thread block.
\par When multi-threads process nested loop join, each thread needs to deal with part of the data. Only when all threads have completed their tasks, the GPU will return results to the CPU. Obviously, the total processing time for tasks on the GPU depends on the last thread. Under this situation, something needs to be done to balance the task for each thread.
\par In this paper, for table T and table S, assuming that the organization of thread blocks in thread grids is $m*n$, the organization of each thread in blocks is $x*y$. So the organization of whole threads in thread grids is $mx*ny$. The number of tuples to be processed in each table by each thread is given by equation~\ref{NBS} and equation~\ref{NBT}:
\begin{equation} \label{NBS}
	\begin{split}
		NB_S=\frac{|S|}{m*x}+1
	\end{split}
\end{equation}
\begin{equation} \label{NBT}
	\begin{split}
		NB_T=\frac{|T|}{n*y}+1
	\end{split}
\end{equation}
\par Suppose that a thread is located in a thread block with the index $(a, b)$, the index of the thread in the thread block is $(c, d)$. The position, where to read $NB_S$ and $NB_T$ tuples from table S and T, is shown in equation~\ref{IDS} and \ref{IDT}:
\begin{equation} \label{IDS}
	\begin{split}
		ID_S=(a*x+c)*NB_S
	\end{split}
\end{equation}
\begin{equation} \label{IDT}
	\begin{split}
		ID_T=(b*y+d)*NB_T
	\end{split}
\end{equation}
\par $ID_S$ and $ID_T$ are the starting positions of the data in two tables. In Figure~\ref{nestedloopfig}, the red box represents the thread, which handles the join operation between  $T1$ and $S3$. In CUDA, the thread block, the index of the thread in the block, sizes of grid and block can be obtained directly, making it easy to connect the data.
\par During thread execution, whenever a match is successful and a new tuple is generated, new tuple needs to be cached, and the result is returned when all threads have finished executing. The biggest problem of multithreading writing data into memory is writing memory conflicts. In order to prevent the conflict, each thread can be allocated a fixed storage space. Whenever a thread generates a result, the results will be written to its corresponding storage space. And when all threads have completed the task, the result will be summed and then returned. At the same time, a local variable is set in each thread to record the number of results generated in the thread.
\par Since the exact number of join results is not known until the join operation is performed, when allocating storage space for each thread, space is set to a maximum possible number of results, that is, $NB_S*NB_T$.
\subsubsection{GPU-Based Hash Join}
\par
\begin{figure}[h]
	\centering\includegraphics[width=0.8\textwidth,height=1in]{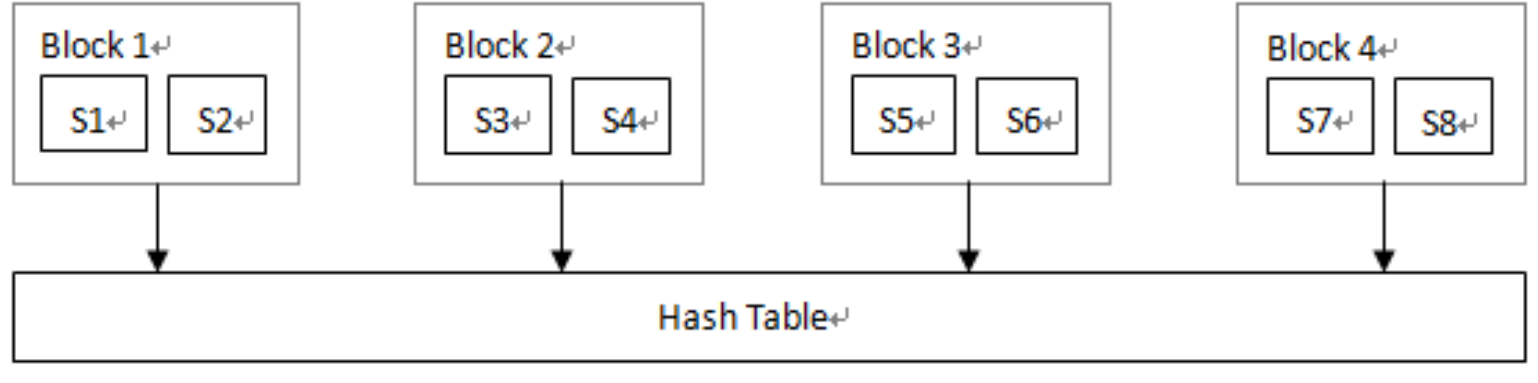}
	\caption{Thread grid organization of hash join}\label{hashjoinfig}
\end{figure}

 According to Section~\ref{nestedloop}, hash join only needs to loop through one data table. So one-dimensional organization of both blocks in grid and threads in the block are more suitable for CUDA. As shown in Figure~\ref{hashjoinfig}, the grid contains four thread blocks, and each thread block contains two threads.
\par Assuming that each thread grid contains $n_{blocks}$ thread blocks, each thread block contains $n_{threads}$ threads, and if the table S is the larger table, the number of tuples in the table S allocated to each thread is given by the equation~\ref{NBSJ}:
\begin{equation} \label{NBSJ}
	\begin{split}
		NB_S=\frac{|S|}{m*x}+1
	\end{split}
\end{equation}
\par In fact, in addition to the last thread in the last thread block, each thread handles $NB_S$ tuples of table S, and the remaining tuples are handled by the last thread. If a block index in the grid is p, and the index of thread in the block is q, the location from where to start reading $NB_S$ tuples in table S can be calculated by equation~\ref{IDSJ}:
\begin{equation} \label{IDSJ}
	\begin{split}
		ID_S=(p*n_{threads}+q)*NB_S
	\end{split}
\end{equation}
\par This paper introduces the hash bucket to handle hash conflicts. Whenever a tuple is detected, it is first positioned to the corresponding hush bucket through the hash function, followed by a small range of detection in the bucket. Hash bucket size is fixed, so the space utilization is slightly inadequate, but the detection efficiency is as good as others.
\par As it shown in Algorithm~\ref{prefilteringalgo}, when the data is sent to the GPU, $NB_T$ and $NB_S$ are sent as well. So $NB_T*NB_S$ is set as the maximum storage space.
\subsubsection{Estimation of Join Results Size}
As mentioned earlier, before the kernel program performs the actual join operation, it is necessary to preallocate enough memory in the CPU for the result. But before the join is completed, we can not know the specific number of results. Usually, memory is set with regard to the size of the Cartesian product. But obviously, when tables are too large, the size of the Cartesian product will be far larger than that of device memory. The existing research is based on the experiments on small data set, so even if the distribution of memory is in accordance with the Cartesian product, it does not exceed the device memory. Although the method proposed in this paper is based on the Cartesian product as well, due to the data pre-filtering and Map-Reduce structure, the performance of the estimation is far better than that of the existing results.
\par Assuming that the ratio of tuples participating in the join of table T and table S is $\alpha$ and $\beta$. So, after pre-filtering, $\beta *|S|$ and $\gamma *|T|$ tuples respectively perform the actual join operation. As mentioned before, tuples with the same $key$ value are passed to the same Reducer. If there are $k$ Reducers, each Reducer deals with $\omega _1*\beta*|S|, \omega _2*\beta*|S| ... ... \omega _k*\beta*|S|$ S-tuples and $\lambda _1*\gamma*|T|, \lambda _2*\gamma*|T| ... \lambda _k*\gamma*|T|$ T-tuples, and the parameters satisfy the following conditions:
\begin{equation} \label{NBSJ}
	\begin{split}
		\omega _1+\omega _2+ ... ... \omega _k=1 \\
		\lambda _1+\lambda _2+ ... ... \lambda _k=1
	\end{split}
\end{equation}
\par $R_{size}$ results need to be allocated storage space:
\begin{equation} \label{rsize}
	\begin{split}
		R_{size}=\gamma *\beta *|S|*|T|\sum_{i=1}^{k}\omega _i*\lambda _i
	\end{split}
\end{equation}
\par Because parameters in equation~\ref{rsize} are all below 1, it will save a lot of space for pre-allocating memory.
\subsubsection{Thread Hierarchy}
In the standard CUDA program, the thread is organized in the form of thread block-thread hierarchy, where the number of thread blocks and the number of threads in each thread grid can be set, but there is only one thread grid. The algorithm in this paper combines Hadoop with the GPU. The Map-Reduce architecture is also a multi-threaded task. Multiple Reducers perform tasks in parallel. Each Reducer corresponds to a thread. Therefore, the hierarchy of the proposed algorithm is thread grid (Reducer)-thread block-thread.
\par In CUDA programming, the number of threads in each block is usually set to a multiple of 32, the number of blocks is based on the amount of data, usually the more the better. Assuming that the GPU has a total of $N$ threads, the reducer number is $n_{reducers}$, and each grid contains $n_{blocks}$ of blocks. Each thread block contains $n_{threads}$ threads. The parameter relationship must satisfy the relationship~\ref{hier}
\begin{equation} \label{hier}
	\begin{split}
		n_{reducers}*n_{threads}*n_{blocks}<N
	\end{split}
\end{equation}
\par That is, in the case where the relationship is established, the bigger product of parameters the better. When the total number of set threads is greater than the number of threads on the GPU, some tasks need to wait for a free thread. In this way, the task waiting and the creation of multiple threads will bring a lot of additional overhead. As shown in Algorithm~\ref{prefilteringalgo}, the number of Reducers can be changed by modifying the parameter $\alpha$.
\subsection{Mixed Programming of GPU and Hadoop}
\par The Map-Reduce program runs on CPU and is based on Java language. While, the GPU kernel program runs on GPU, based on the CUDA programming model and the C language. Therefore, the combination of Hadoop and GPU involves a compatibility  problem between Java and C. This experiment is programmed through the JNI interface to complete this work. Since the file format of the CUDA program is .cu, and the traditional C language is the .c file, there is a need of an additional .c file as a middleware. Through this middleware, GPU kernel function can be called. And then Map-Reduce will call the middleware, so as to achieve the join of CPU and GPU. Because the memory cannot be shared between CPU and GPU, the same data from Java to C needs a copy, and then another copy from C memory space to GPU device memory. That is, to transfer data from Map-Reduce to GPU, and a total of two copies are required. Obviously, this introduces additional overhead and memory footprint, so future research should continue to solve this problem.
\section{GPU-Based Theta Join with Hadoop}
\par Given the importance of Hadoop, there are many research focused on using Map-Reduce to deal with theta join operations. Koumarelas\cite{Koumarelas}, Okcan\cite{Okcan} and Penar\cite{Penar} achieved theta join operations of two tables on Map-Reduce, meanwhile Yan\cite{Yan}, Zhang\cite{Zhang} and Changchun\cite{Changchun1, Changchun2} achieved a multi-table theta join. Augustyn\cite{Augustyn} used the GPU to estimate the join selectivity for theta join of two tables.
\par This section focuses on how to combine Hadoop with a GPU to handle theta join operations based on two data tables, multi-table join will be studied later.
\subsection{Classic Theta Join Algorithm on Hadoop}
\par
\begin{figure}[h]
	\centering\includegraphics[width=1.5in]{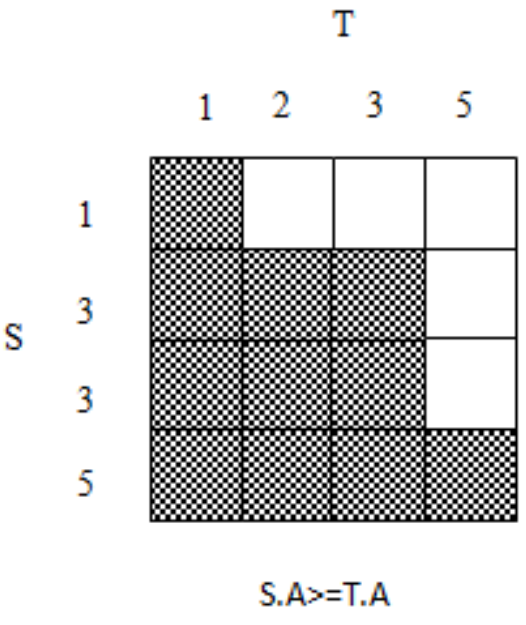}
	\caption{Matrix in classic algorithm}\label{nonequifig}
\end{figure}

 For theta joins, simply passing tuples with the same join key to the same Reducer is not sufficient to get the full result. Because in theta join a tuple does not only need to match its tuples with the same join keys, but also need to match the tuples that are greater or less than the join key. As shown in Figure~\ref{nonequifig}, the areas represented by the shaded parts all need to be matched.
\par
\begin{figure}[H]
	\centering
	\subfigure[Classic algorithm]{
		\label{classa}
		\includegraphics[width=1.5in]{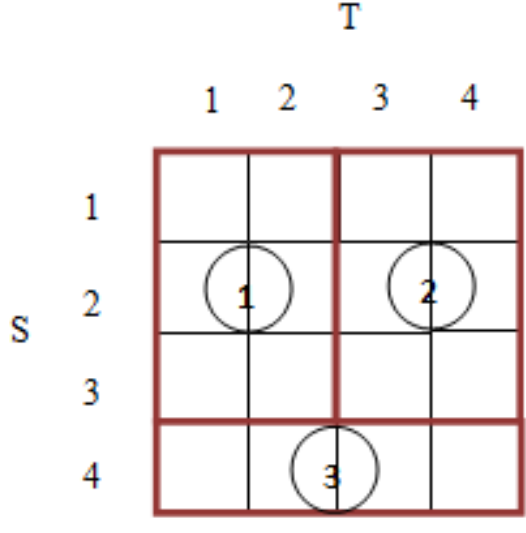}}
	\subfigure[Developed algorithm]{
		\label{classb}
		\includegraphics[width=1.5in]{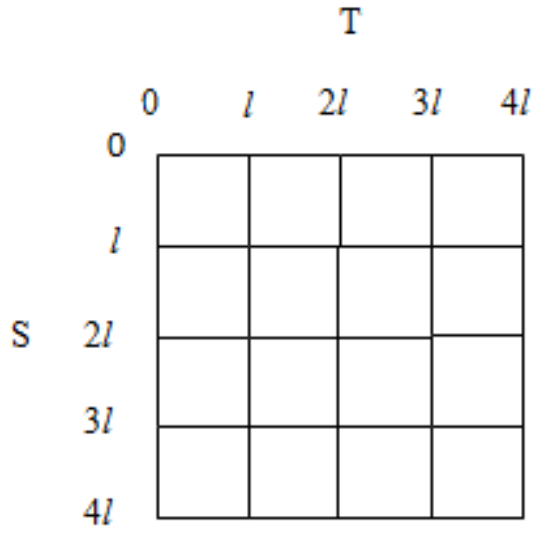}}
	\caption{Theta join algorithm on Map-Reduce}
	\label{class}
\end{figure}

 Okcan\cite{Okcan} proposed a classic Map-Reduce processing theta join operation algorithm, named M-Bucket-I. As shown in Figure~\ref{classa}, the Cartesian product of two tables is represented by a two-dimensional matrix. The matrix is divided into many areas, each region corresponds to a Reducer, and each region has an ID. In Map function, for each tuple, a $(ID, <Record, tag>)$ key-value pair is generated for each region intersecting it, where $tag$ is the identity that identifies the tuple source. As a result, tuples with the same ID, that is, tuples located in the same area, are passed to the same Reducer for processing. In this way, it is possible to ensure that each element in the matrix can be processed and led to the correct result.
\par In this algorithm, only the area where the result is determined is processed, and no actual processing is performed on the data that does not produce the result area. However, this algorithm requires some information about the original data in advance, which means additional overhead is needed.\begin{figure}[H]
	\centering
	\subfigure[Greater and greater or equal]{
		\label{greena}
		\includegraphics[width=1.5in]{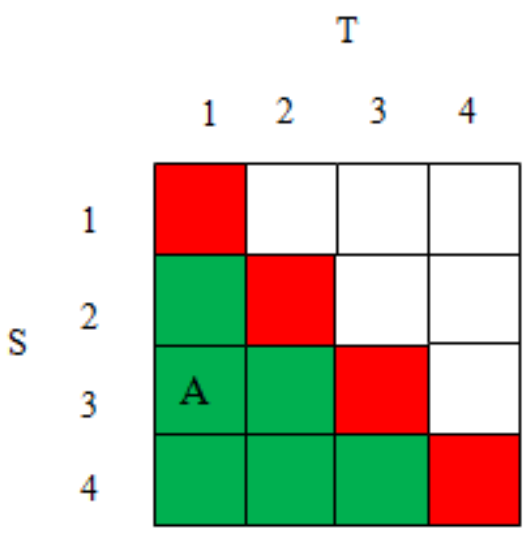}}
	\subfigure[Less and less or equal]{
		\label{greenb}
		\includegraphics[width=1.5in]{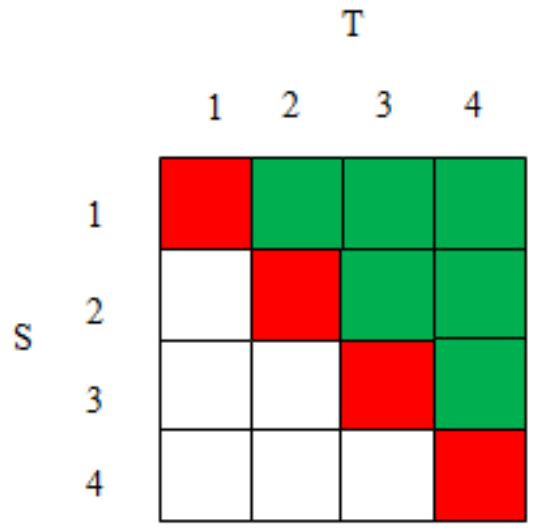}}
	\subfigure[Not equal]{
		\label{greenc}
		\includegraphics[width=1.5in]{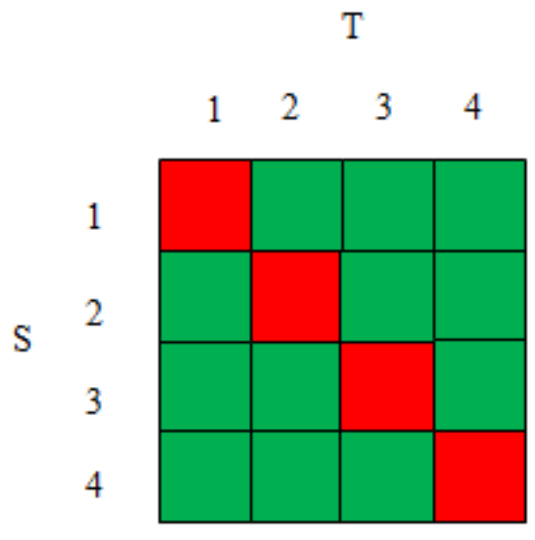}}
	\caption{Organization of matrix under different join conditions}
	\label{class}
\end{figure}
\subsection{Improved Theta Join Algorithm on Hadoop and GPU}
\par To solve this problem, this paper presents an improved algorithm. On the basis of the classic algorithm, this algorithm can further identify certain areas, where tuples from two tables satisfy the join condition.
\par Similar to the  M-Bucket-I, this algorithm calculates the range of each table join key and k quantiles of each range. Figure~\ref{classb} shows the organization of the two-dimensional matrix when k is 4. In this matrix, each element does not represent a tuple pair, but a range pair. Unlike the  M-Bucket-I where the divided area is the rectangle of any size, this algorithm divides the area into strict squares. The advantage of this division is, for some regions, if any one of the tuple pairs meets the join condition, there is no need to judge the data in this region for the join.

\par As shown in Figure~\ref{greena}, when the join condition is that join key attribute value of S-table is greater than or greater than or equal to that of T-table, any pair of tuple pairs in the green region satisfies the join condition, and the tuple pair in the red region needs to be judged. While the tuple pair in the white region must not satisfy the join condition. Figure~\ref{greenb} shows that of S-table is less than or less than or equal to that of T-table. And Figure~\ref{greenc} show the condition that two values are not equal. So only the red area of the data needs to be sent to the GPU for theta join processing. And the other areas can be judged directly by the area ID. The pseudo code is shown in algorithm~\ref{mapreduce}.
\begin{algorithm}[ht]
	\scriptsize
	\caption{theta join on Map-Reduce}
	\label{mapreduce}
	\begin{algorithmic}[1]
		\STATE Map($key$:null, $value$: a tuple from a split of either table)
		\STATE $join\_key \leftarrow$ extract the join key from $value$;
		\IF{$value\in S$}
		\STATE $RegionID.x \leftarrow$ bucket ID of $value$ according to its join key;
		\FOR {all possible $RegionID.y$ according to the join condition}
		\STATE emit($<RegionID.x, RegionID.y >$$,<value, "S">$);
		\ENDFOR
		\ENDIF
		\IF{$value\in T$}
		\STATE $RegionID.y \leftarrow$ bucket ID of $value$ according to its join key;
		\FOR {all possible $RegionID.x$ according to the join condition}
		\STATE emit($<RegionID.x, RegionID.y >$$,<value, "T">$);
		\ENDFOR
		\ENDIF
		\STATE Reduce($key^,$:$RegionID$, $value\_list$: tagged tuples corresponding in $RegionID$)
		\STATE $T$ $\leftarrow$ null;
		\STATE $S$ $\leftarrow$ null;
		\FOR {each $tuple$ t in $value\_list$}
		\STATE add t to S or T according to its tag;
		\ENDFOR
		\IF{$RegionID.x$ == $RegionID.y$}
		\STATE $result \leftarrow$ $GPU\_theta\_join(S, T)$;
		\FOR {each $record$ in $result$}
		\STATE emit(null, $record$);
		\ENDFOR
		\ELSIF{$RegionID.x$ matches $RegionID.y$ according to the join condition}
		\STATE $cartesian\_result \leftarrow$ do cross join for $S$ and $T$;
		\FOR {each $record$ in $cartesian\_result$}
		\STATE emit(null, $record$);
		\ENDFOR
		\ENDIF
	\end{algorithmic}
\end{algorithm}
\par As shown in Figure~\ref{greena}, the A region coordinates are (3, 1) and the abscissa is greater than the ordinate. Therefore, any tuple in the A region satisfies the condition, that join key attribute value of table S is greater than that of table T. When the abscissa is equal to the ordinate, the region may have a join result, so it is necessary to send it to the GPU for theta join judgment.
\par In the theta join operation on the GPU, nested loop join algorithm can be used, the thread organization form and the processing mode are the same as those of the GPU in Chapter~\ref{gpubej}.
\par The method used in this article only applies the data in the diagonal area to the GPU, so it is only necessary to allocate storage space for the data in the diagonal area, which can increase the storage space utilization.

\section{Experimental Results}
\par This section compares the methods proposed in this paper, through the existing GPU accelerated join operation algorithm to verify whether it has improved performance, being more efficient. At the same time, compared with the CPU implementation of the proposed algorithm in this paper, to verify whether the GPU implementation has the speedup for the join operation.
\par The experiments done in this section are based on larger datasets. Unless special instructions, all experimental raw data are TPC-H data sets. GPU devices running on the Linux operating system, the version of Ubuntu 14.04, and Hadoop is version 2.6.0.
\subsection{Nested Loop Join}
\par This section focuses on the experiments of the nested loop join. It is compared separately with the GPU accelerated nested loop join with a single device and the proposed algorithm in CPU. At the same time by changing the value of $\alpha$ in algorithm~\ref{prefilteringalgo} to change the number of Reducers started by Map-Reduce, changes in the execution time can be observed. At the same time, through the synthetic data, the performance of the method can be observed. If there is no special description, the value of $\alpha$ is 100. This experiment is based on the small dataset, because the cost of the nested loop join is very large, the traditional GPU acceleration method is not able to effectively support large data sets.
\subsubsection{Comparison of Nested Loop Join with Single GPU}
\par In this experiment, in order to ensure the reliability of the experimental results, the proposed method is also implemented on a single GPU device, and Hadoop is a pseudo-distributed structure. As shown in Figure~\ref{gpunested}, when the data set is small, the efficiency of the proposed method is lower than that of the traditional one. However, the execution time of the traditional method is significantly increased with the increase of the original dataset, which is different to the proposed algorithm. As shown in the figure, it will have at least one times the speedup over the traditional one, which means under the same accelerating condition of GPU, the proposed method is more efficient.
\begin{figure}[h]
	\centering\includegraphics[width=3in]{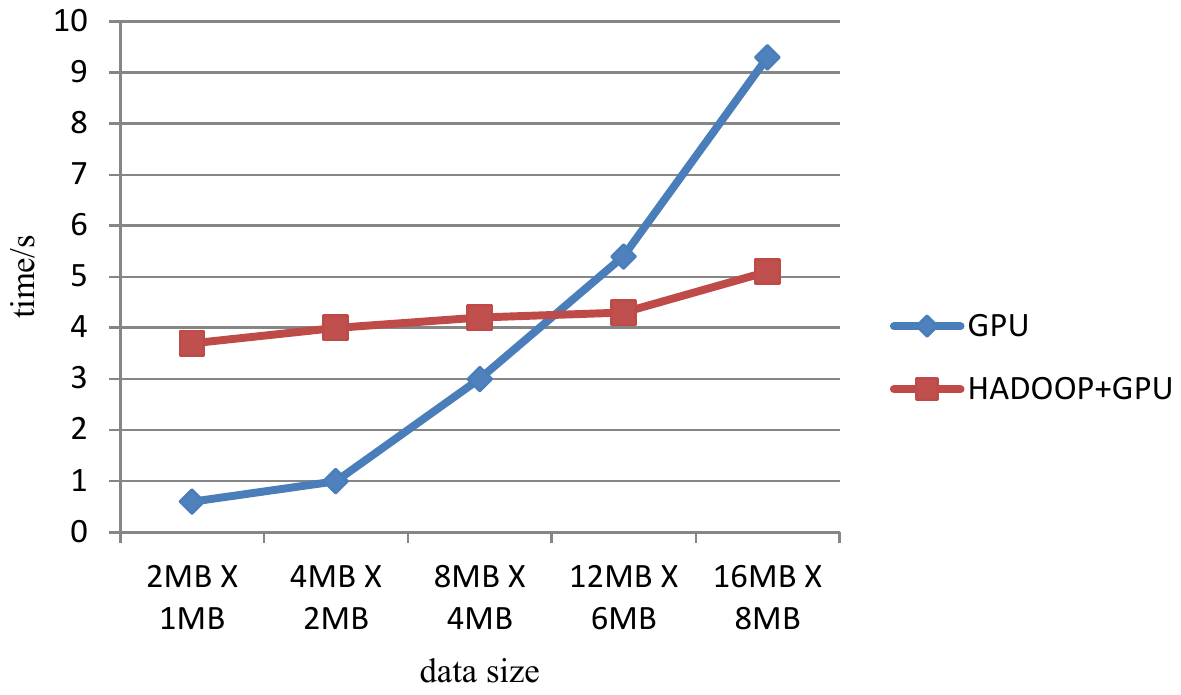}
	\caption{Comparison of nested loop join with single GPU}\label{gpunested}
\end{figure}
\par Data that does not appear in the final result is filtered out, and the data that actually performs the join operation is much smaller than the original data set. After pre-filtration, the actual processing time is greatly reduced, but the pre-filtration process also introduces additional time overhead.
\par Therefore, when the data set is small, the gain effect of the pre-filtering is lower than that of the additional cost, so the method proposed in this paper has a long execution time at first. With the increase of the data set, the gain effect exceeds the overhead, so the method proposed in this paper is significantly lower than the traditional method.
\subsubsection{Comparison of Nested Loop Join with CPU on TPC-H Data Set}
\par
\begin{figure}[h]
	\centering\includegraphics[width=3in]{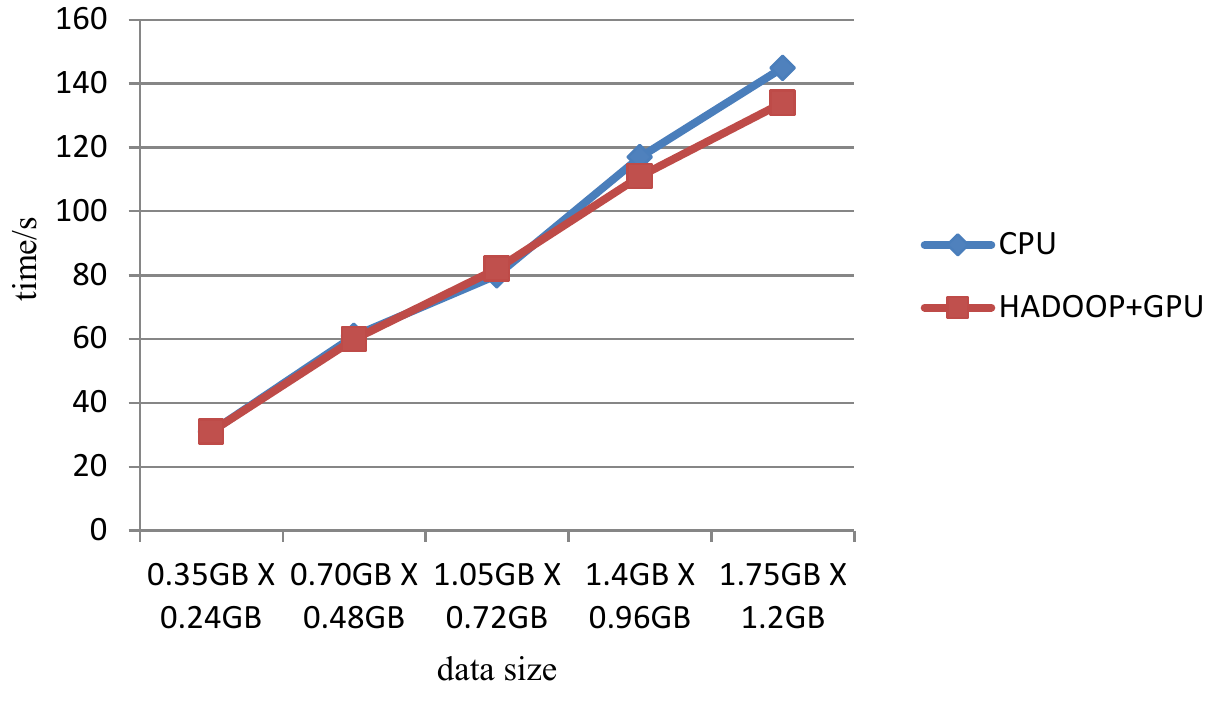}
	\caption{Comparison of nested loop join with CPU on TPC-H data set}\label{cpunested}
\end{figure}

 The value of $\alpha$ in the CPU is 1, which ensures the best execution efficiency. Except for the value of $\alpha$ and the equipment, the implementation process of GPA and CPU is exactly the same. As shown in Figure~\ref{cpunested}, compared with the CPU implementation, using the GPU to achieve the proposed nested loop join algorithm has no obvious speedup. This is because, in Reduce phase of the second round of Map-Reduce of CPU version, only tuples with the same join key will be passed to the same Reducer. In this experiment, the distribution of the TPC-H data join keys is even. Each join key value corresponds to only a small number of tuples. Therefore, the actual execution time of the join operation in the CPU version is very small, most of the time is taken by data pre-filtering. And GPU can only accelerate small proportion of data, so the effect is not obvious.
\subsubsection{Comparison of Nested Loop Join with CPU on Synthetic Data Set}
\par
\begin{figure}[h]
	\centering\includegraphics[width=3in]{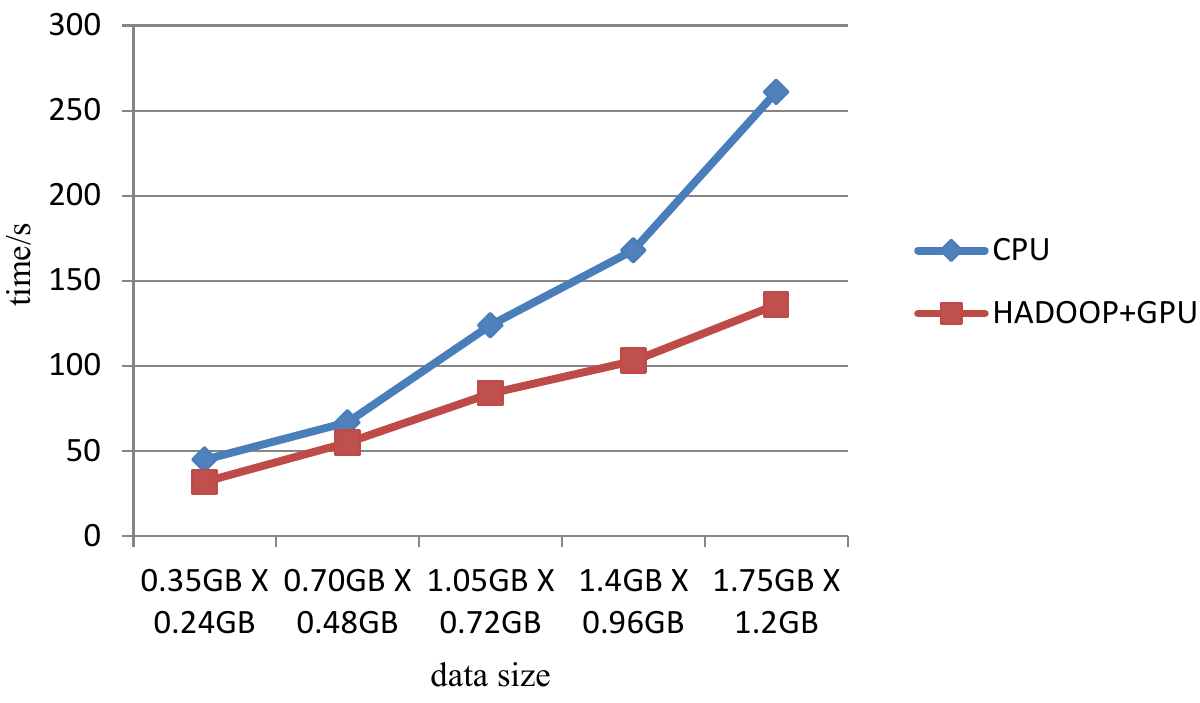}
	\caption{Comparison of nested loop join with CPU on synthetic data set}\label{mannested}
\end{figure}

 Based on the unobvious effect of the proposed algorithm on the TPC-H data set, the synthetic data set is made for another experiment. In the synthetic data set, the number of tuples corresponding to each join key is more than that of the TPC-H data set. To ensure the reliability of the experimental results, the distribution of join keys has been modified on the basis of the experimental data set shown in Figure~\ref{cpunested}. Specifically, dividing the join key by a positive integer, the number of tuples corresponding to each join key increases. As can be seen from Figure~\ref{mannested}, in the synthetic dataset, compared to the CPU, using the GPU can get nearly 2 times the speedup. Through Figure~\ref{gpunested} and Figure~\ref{cpunested}, the proposed method can be applied to larger data sets than traditional GPU algorithms. Because the join key value range in the synthetic data set is much smaller than that in the TPC-H dataset, when the value of $\alpha$ is small, a good speedup can be obtained. Therefore, $\alpha$ is 5 in this experiment.
\subsubsection{Effect of $\alpha$ on the Execution Efficiency of Nested Loop Join}
\par
\begin{figure}[h]
	\centering\includegraphics[width=3in]{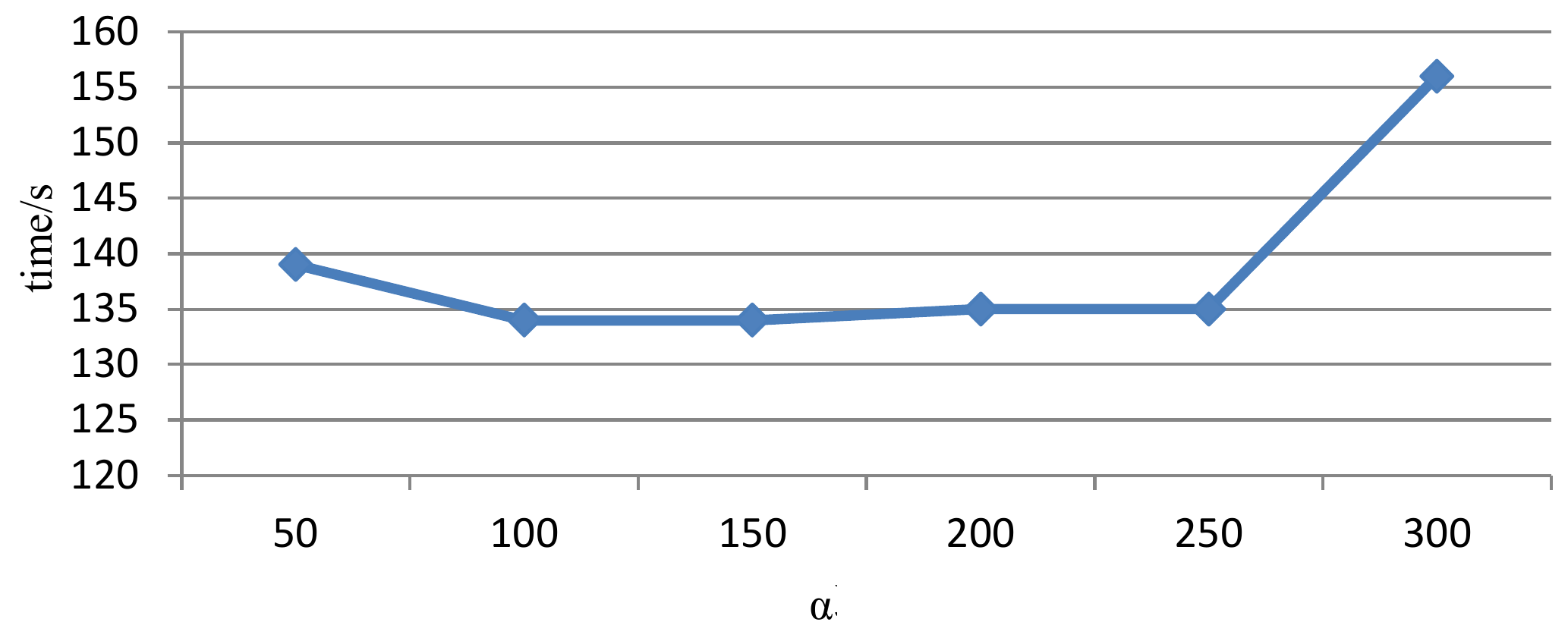}
	\caption{Effect of $\alpha$ on the execution efficiency of nested loop join}\label{nesteda}
\end{figure}

 In the algorithm~ref{prefilteringalgo}, the value of $\alpha$ will affect the parallel number of the algorithm. To clarify the influence of $\alpha$ on the performance of the algorithm, we can observe the execution time of the algorithm by changing the value of $\alpha$. This experiment uses a TPC-H data set, the data set size is 1.75GB and 1.2GB. The experimental results are shown in Figure~\ref{nesteda}. When $\alpha$ is small, with the increase of $\alpha$, the execution time of the algorithm decreases gradually. When $\alpha$ increases to a certain extent, with the increase of $\alpha$ value, time also increases.
\par When the value of $\alpha$ is small, the number of Reducers is larger, resulting in the total number of threads of the algorithm beyond the total number of GPU devices. Some tasks need to wait for the other tasks to be completed before they can be assigned to the free thread to perform the task. The waiting time plus the overhead of creating multiple threads make the execution time of the algorithm higher when the value of $\alpha$ is less. When the value of $\alpha$ is large enough that the number of total threads of the algorithm is smaller than that of the total number of GPU devices, with the increase of $\alpha$, the number of Reducer decreases. It leads to the decrease of the total number of threads, the increase of data amount processed by each thread, so the algorithm execution time will increase with the increase of $\alpha$. It can be seen from Figure~\ref{nesteda} that the algorithm will obtain the ideal execution effect when the value of $\alpha$ is between 100 and 250.
\subsection{Hash Join}
\par This section focuses on the experiment of accelerating the hash join algorithm. The experimental results are observed by comparing the hash algorithm with the single GPU and the CPU implementation. And in the synthetic data set, the efficiency and the speedup of the proposed method can be observed. Additionally, by changing the $\alpha$ value, this section observes the influence of $\alpha$ on the algorithm. If there is no special instruction, $\alpha$ value is 10000.
\subsubsection{Comparison of Hash Join with Single GPU}
\par In this experiment, the value of $\alpha$ is 10000, which is much larger than the $\alpha$ when dealing with nested loop join. This is because the hash join calculation task is much smaller than the nested loop join, so larger amounts of data processed in each Reducer can give full play to the advantages of GPU.
\par
\begin{figure}[h]
	\centering\includegraphics[width=3in]{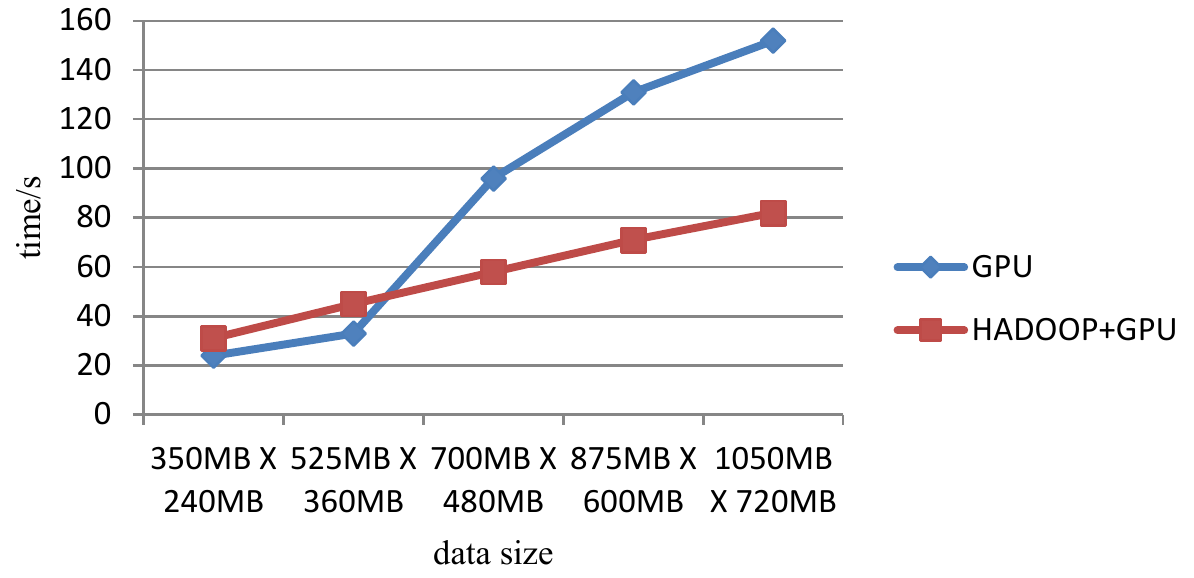}
	\caption{Comparison of hash join with single GPU}\label{gpuhash}
\end{figure}

 As can be seen from the Figure~\ref{gpuhash}, when the data set is small, the proposed method has lower efficiency, compared to the traditional GPU algorithm. When the data set size is good enough, the proposed method is better than the traditional GPU algorithm. The experimental results show that the proposed method can achieve 2 times the speedup over traditional one, which means under the same accelerating condition of GPU, the proposed method is more efficient.
\subsubsection{Comparison of Hash Join with CPU on TPC-H Data Set}
\par When the data set is small, most of the time is used to process the data pre-filter. When the data pre-filter takes longer than the time it reduces on the actual join operation, the total time will increase. While as the data set continues to increase, the method proposed in this paper will achieve better results, because the data pre-filter is less time-consuming. By comparing with Figure~\ref{gpunested}, it is found that traditional GPU processing algorithms can handle larger amounts of data when handling hash joins. This is because when the data set is large, the hash join algorithm performance is much higher than the nested loop join, so it can handle more data sets.
\begin{figure}[h]
	\centering\includegraphics[width=3in]{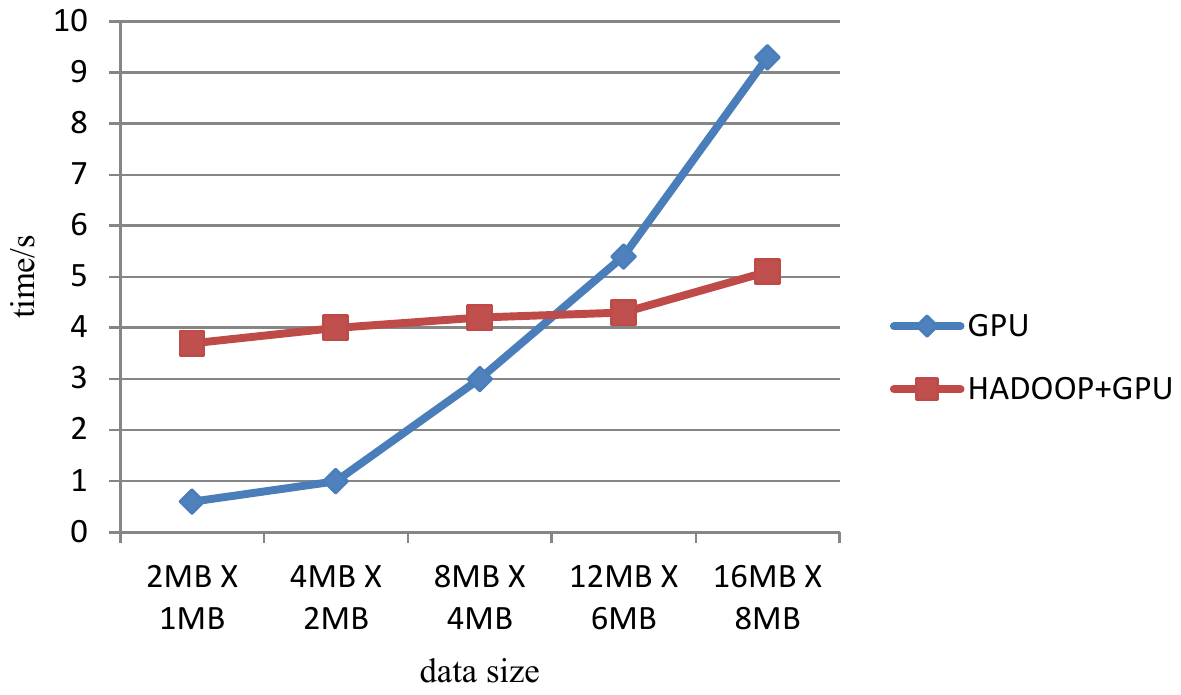}
	\caption{Comparison of hash join with CPU on TPC-H data set}\label{gpunested}
\end{figure}
\subsubsection{Comparison of Hash Join with CPU on Synthetic Data Set}
\par
\begin{figure}[h]
	\centering\includegraphics[width=3in]{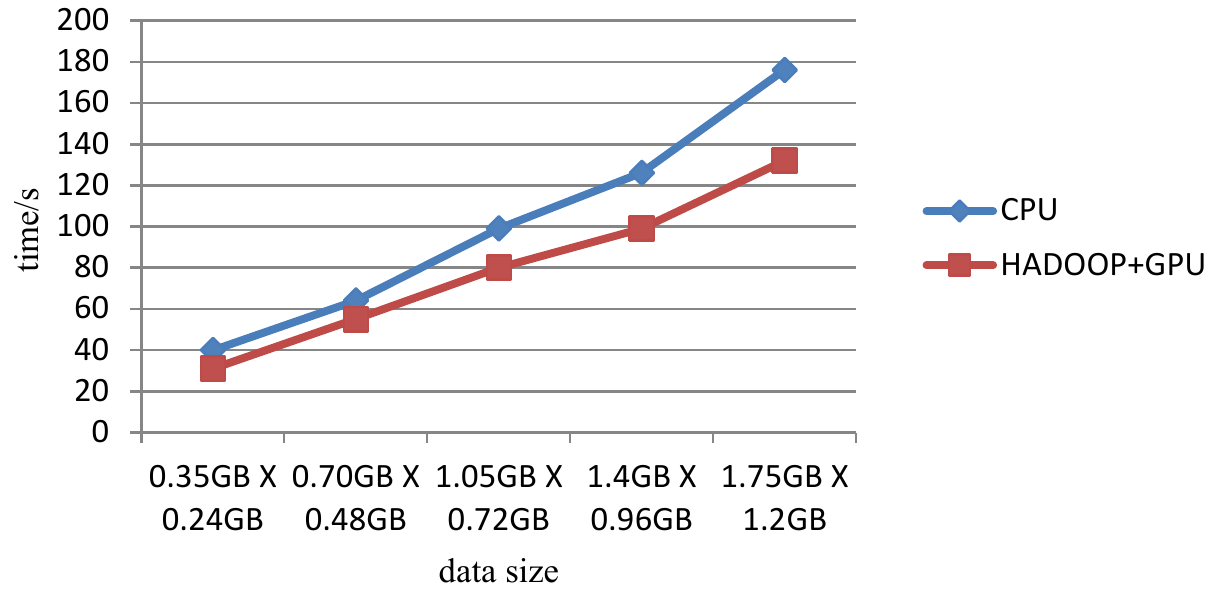}
	\caption{Comparison of hash join with CPU on synthetic data set}\label{manhash}
\end{figure}

 Data is synthesized in the same way as nested loops. As shown in Figure~\ref{manhash}, in the synthetic data, the proposed GPU algorithm is better than its CPU version, with 1.3 times speedup. Hash join speedup is not good as nested loop join, because the nested loop join spends more time in the actual join processing, so it can be optimized in the proportion of large.
\subsubsection{Effect of $\alpha$ on Execution Efficiency of Hash Join}
\par
\begin{figure}[h]
	\centering\includegraphics[width=3in]{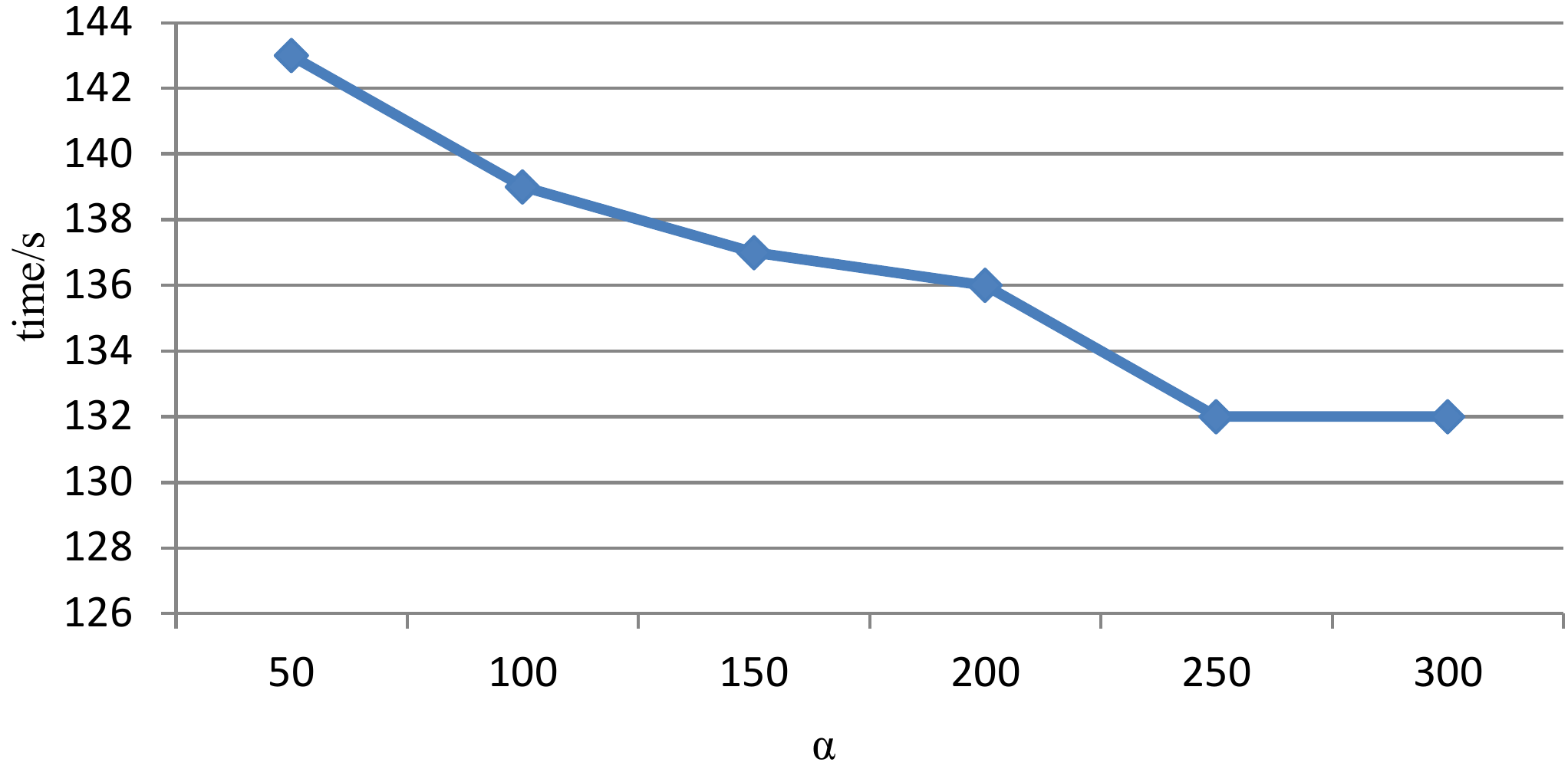}
	\caption{Effect of $\alpha$ on the execution efficiency of hash join}\label{hasha}
\end{figure}

 This experiment uses TPC-H data set, the data set size is 1.75GB and 1.2GB. Similarly, $\alpha$ is changed in the experiment to observe its effect on algorithm performance. Figure~\ref{hasha} shows the change in the execution time of the hash join algorithm proposed in this paper when the value of $\alpha$ changes. When $\alpha$ is small, the execution time of the algorithm decreases obviously as $\alpha$ becomes larger. When the value of $\alpha$ is too small, the total number of threads of the algorithm exceeds the total number of threads of the GPU device, so the execution time is longer. Since the computational task of the hash join is much smaller than the nested loop join, the overall execution time does not increase significantly when the $\alpha$ value continues to increase, although the amount of data processed in each Reducer increases. If the value of $\alpha$ continues to increase to a certain extent, the algorithm's overall execution time will continue to increase.
\subsection{Theta Join}
\par This section compares the theta joins. Because there is no available mature GPU accelerating theta join, this experiment compares with the CPU. This experiment is based on a small dataset. In order to ensure the validity of the experimental results, implementation details are exactly as described in Algorithm~\ref{mapreduce}
\par
\begin{figure}[h]
	\centering\includegraphics[width=3in]{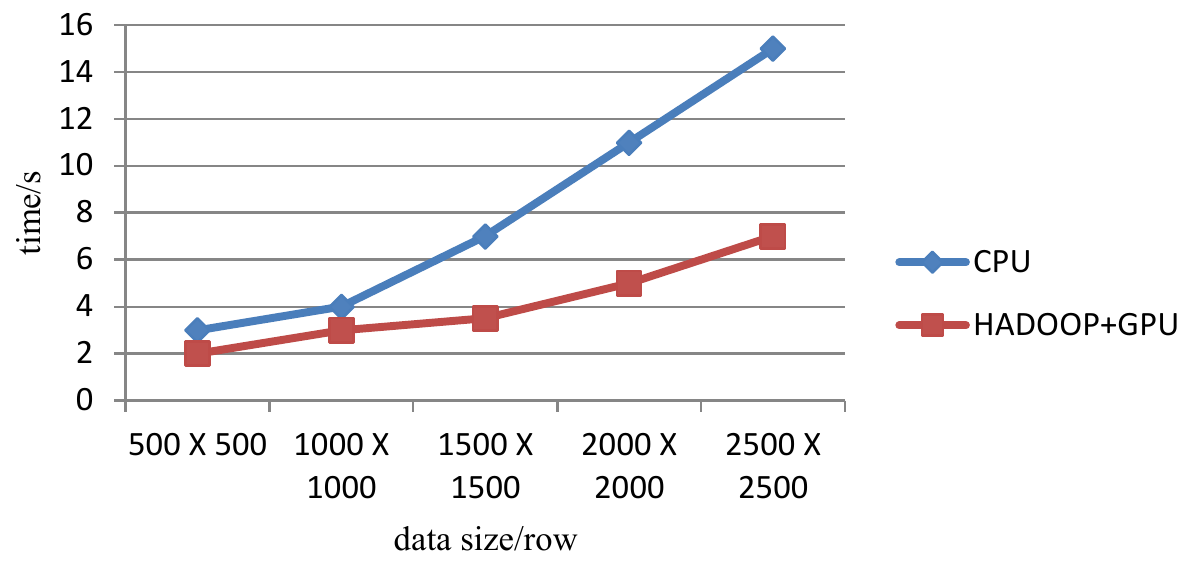}
	\caption{Comparison of theta join on small data set}\label{resultnon}
\end{figure}

 In contrast to the equi join, the theta join has a high selectivity, so the size of the theta join results is larger than the magnitude of input data. The results of the experiment are shown in Figure~\ref{resultnon}. It can be seen from the figure that the Hadoop-based theta join algorithm on the GPU is better than its CPU version, and the speedup is about twice as much as the latter, which means under the same accelerating condition of GPU, the proposed method is more efficient.
\par
\begin{figure}[h]
	\centering\includegraphics[width=3in]{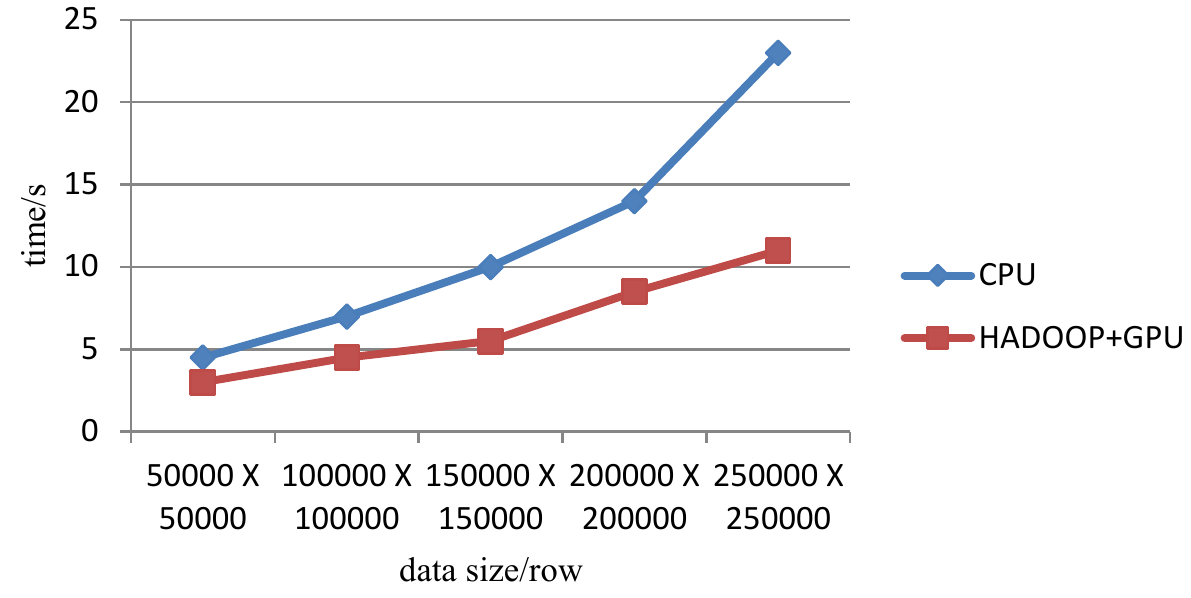}
	\caption{Comparison of theta join on small data set}\label{nonresultnon}
\end{figure}

 We also experimented on larger datasets. Since the join results are out of memory when the data set is large, we do not store the results. As shown in Figure~\ref{nonresultnon}, when the data set is large, GPU implementation is still better than the CPU implementation.
\section{Conclusion}
\par This paper focuses on Hadoop-based join operation acceleration tasks on image processors (GPUs). GPU was originally developed as an image processing, and nowadays, more and more GPU applications appear in general computing tasks, such as machine learning, data mining and other fields. Based on the GPU's powerful computing power and high parallelism, there is a lot of research focused on using it to speed up database operations. In the field of the modern database, the join operation as a computationally intensive task is the main problem.
\par In research results of existing GPU accelerating join operations, although the use of its strong parallelism significantly increases the efficiency of join execution, it is not good because of the limited storage resources of the GPU device and the limited functionality of the universal programming language CUDA. The existing research results are based on smaller data sets, and therefore can not be applied to the practical application on a large scale.
\par Based on this idea, the distributed computing platform Hadoop is combined with the GPU, by referring to the idea of a CPU filtering join algorithm. By initially filtering the raw data table through the first round of the Map-Reduce task at the CPU, it will filter the tuple that does not appear in the results, and only send the connectable tuples to the GPU device for actual join operations. By reducing the amount of data actually processed, it is possible to reduce the utilization of the storage space on the device while improving the efficiency of the algorithm execution so that it can handle a larger amount of data. At the same time, it is possible to estimate the number of join results more accurately without introducing additional overhead, to allocate accurate storage in advance and reduce the storage space occupancy rate. In addition to the equi join, this article is the first to use the GPU to accelerate the theta join operation, which still uses the combination of Hadoop and GPU.
\par The followings are the conclusions:
\begin{itemize}
	\item The proposed algorithm is more efficient than the existing single GPU device, and it can be applied to the larger dataset. Compared with the CPU implementation, the GPU algorithm proposed in this paper has no obvious speedup on the dataset with fewer key numbers and fewer tuples corresponding to each join key. However, the algorithm proposed in this paper can achieve 2 times the speedup on a dataset with more corresponding tuples.
	\item The accelerating hash join algorithm proposed in this paper can achieve 2 times the speedup, compared with the existing GPU acceleration hash join. Similarly, the GPU implementation of the algorithm proposed in this paper has no obvious speedup compared to the CPU implementation, when the number of connected keys is large, but each join key value corresponds to a few tuples. When the number of corresponding tuples is big enough, GPU implementation can get 1.3 times the speedup.
	\item In this paper, the theta join processing algorithm, compared to the CPU implementation, GPU implementation can get 2 times the speedup.
\end{itemize}
Compared with the existing research results, the research content of this paper has achieved the following innovative achievements:
\begin{itemize}
	\item This article is the first to use the filter join algorithm on the GPU to deal with the equi join operation. One round of Map-Reduce filters out non-connectable tuples and sends only connectable tuples to the GPU device. By reducing the processing time of the actual join operation and the occupancy rate of the device memory, it can handle more data.
	\item The proposed method can accurately estimate the size of the equi join result without introducing additional overhead, and allocate the appropriate storage space for the result, making the GPU storage space more efficient, more suitable for large-scale data sets.
	\item Among the existing GPU accelerating equi join operation, this article is the first to have an experiment on a larger data set (GB level).
	\item This article is the first to use GPU to accelerate theta join operations.
\end{itemize}
\par Although there are a lot of research results on GPU accelerating join, these results are not enough to be applied to the commercial database system. So the future work should continue to conduct in-depth research, including the following:
\begin{itemize}
	\item The future work should use multiple GPU devices to deal with large datasets on the join operation, in a distributed architecture. Although in this article Hadoop and GPU were combined, because of limited resources, the experiment is not completed in the real large data set. Future research should deal with TB-level datasets on multiple GPU devices.
	\item This paper only implements the join operation of two tables. Future research work should include more complicated join operations, such as accelerating multi-table join and similarity join.
\end{itemize}

\paragraph{Acknowledgement}This paper was partially supported by NSFC grant U1509216,61472099, National Sci-Tech Support Plan 2015BAH10F01, the Scientific Research Foundation for the Returned Overseas Chinese Scholars of Heilongjiang Province LC2016026 and MOE-Microsoft Key Laboratory of Natural Language Processing and Speech, Harbin Institute of Technology.

\section{References}

\bibliographystyle{plain}
\bibliography{Join}

\end{document}